\documentclass{aa}  

\usepackage{graphicx}
\usepackage{tabularx}
\usepackage{amsmath}
\usepackage[colorlinks=true,linkcolor=magenta,citecolor=blue,filecolor=cyan,final=true]{hyperref}
\usepackage{txfonts}
\usepackage{multirow}
\newcommand{\CellWithForceBreak}[2][c]{
\begin{tabular}[#1]{@{}c@{}}#2\end{tabular}}

\begin{document} 

    \title{Tracing magnetic flux rope footpoints in AR~12975 with coronal dimmings and data-driven magnetofrictional simulations}

   \author{A. Wagner
          \inst{1}\thanks{Shared first authors}
        \and     
        A. Razquin \inst{2}\footnotemark[1]
        \and
        A.M. Veronig\inst{2, 3}
        \and
        E. K. J. Kilpua\inst{1}
        \and
        K. Dissauer \inst{2,4}
        \and
        J. Pomoell \inst{1}
}

    \institute{Department of Physics, University of Helsinki, P.O. Box 64, FI-00014, Helsinki, Finland\\ 
    \email{andreas.wagner@helsinki.fi}
    \and
    University of Graz, Institute of Physics, Universitätsplatz 5, 8010 Graz, Austria\\
    \email{amaia.razquin-lizarraga@uni-graz.at}
    \and 
    University of Graz, Kanzelh\"ohe Observatory for Solar and Environmental Research, Kanzelh\"ohe 19, 9521 Treffen, Austria
    \and
    NorthWest Research Associates, 3380 Mitchell Lane, Boulder, CO 80301, USA
    }
    \date{Received: 3 July 2026; Accepted: 15 September 2026}

\abstract
{Coronal mass ejections (CMEs) are driven by the eruption of magnetic flux ropes (MFRs), whose footpoints are often observed indirectly as coronal dimming regions. However, the relationship between observed dimmings and true MFR footpoint locations remains uncertain due to limitations and uncertainties in both detection methods as well as coronal magnetic field modelling caused by the complexity of eruptive events.}
{We aim to investigate the M4 flare and CME event on 28 March 2022 in AR~12975 to determine how well observed coronal dimmings trace the footpoints of an erupting MFR. Furthermore, we intend to assess how these signatures compare with the MFR evolution from a data-driven magnetofrictional (TMFM) simulation.}
{We combined multiwavelength SDO/AIA observations with a time-dependent data-driven magnetofrictional simulation of the active region (AR). We identified coronal dimmings using intensity-thresholding and minimum-intensity maps. For challenging regions we used a tailored region growth algorithm from strong gradients. We extracted MFR structures in the simulation using a combined twist and squashing factor metric. We projected MFR footpoints onto photospheric magnetograms for direct comparison and derived magnetic fluxes from both dimmings and the modelled MFR.}
{The simulation reproduces the formation, evolution, and eruption of the observed MFR whose footpoints migrate during the eruption. Moreover, the footpoints match well with the dimming location. The southern dimming shows characteristics of a moving flux rope dimming, while the northern dimming behaves as a shrinking flux rope dimming that is rapidly closed by the flare ribbons. Quantitative comparisons of area and magnetic flux show good overall agreement between simulated footpoints and observed dimmings, with discrepancies mainly arising from limitations of the detection methods and the inherent complexity of the event. The dimmings at the northern footpoint region is particularly challenging in observations due to pre-existing low intensities and its fast shrinkage.} 
{}

   \authorrunning{A.~Wagner \& A.~Razquin, et al.} 
   \titlerunning{Simulated flux rope footpoints compared to coronal dimmings}
   \maketitle

\section{Introduction}
\label{Sect: intro}

The main structures involved in solar eruptions are magnetic flux ropes (MFRs), which are coherent twisted magnetic structures whose field-lines wind around a common axis \citep{Chen2017, Green2018}. MFRs may either pre-exist in the corona or form through magnetic reconnection during the eruption process \citep{chen2011coronal, patsourakos2020decoding}. Once an MFR becomes unstable, it expands and leads to a coronal mass ejection (CME), i.e. a large expulsion of magnetised plasma into interplanetary space \citep{Webb2012}. As the MFR erupts, coronal plasma is depleted from its footpoints, producing localised and pronounced decreases in the intensity of soft X-ray (SXR) and extreme-ultraviolet (EUV) emission in conjugated magnetic polarities \citep{sterling1997yohkoh, thompson1998soho, kahler2001origin, attrill2006using, miklenic2011coronal, pan2021preeruption}. At the same time, the overlying magnetic field is stretched and may partially open, leading to widespread regions of reduced coronal emission that are generally characterised by a smaller drop in intensity \citep{mandrini2007cme, Dissauer2018b}. These sudden reductions in intensity are known as coronal dimmings and are interpreted to be signatures of plasma evacuation associated with CME expansion \citep{hudson1996long, sterling1997yohkoh, Zarro1999soho, harra2001material, Jin2009coronal, lopez2017mass, vanninathan2018plasma, Veronig2019}. 

Localised coronal dimming regions rooted at the footpoints of the erupting MFR have been traditionally referred to as core dimmings, while the extended and shallower dimming regions have been classified as secondary dimmings \citep{thompson2000,mandrini2007cme,Dissauer2018a}. However, this distinction provides limited physical insight into the eruption process and the mechanisms responsible for the observed dimming regions. To address this limitation, \citet{veronig2025coronal} proposed a new framework for characterising coronal dimmings that accounts for the magnetic flux systems and (reconnection) processes involved in the dimming signature by relating the dimmings to the flare ribbon evolution. Within this framework, structures previously grouped as core dimmings are sub-divided into different categories of flux rope dimmings, which may be stationary, shrinking, or moving, depending on the evolution of the erupting MFR and the associated newly reconnected magnetic field lines. 

There are still relatively few examples in the literature demonstrating how  the observed dimming signatures should be interpreted within the context of the new classification scheme \citep{purkhart2025magnetic, razquin2026magnetic}. Moreover, different dimming detection methods may overlook certain dimming regions or merge them together. Coronal dimming detection is most often performed using thresholding techniques applied to base-difference \citep{thompson1998soho, attrill2008recovery,  podladchikova2005nemo, bewsher2008relationship, aschwanden2016global}, logarithmic base-ratio \citep{Dissauer2018b, chikunova2020coronal, jain2024estimating, razquin2025coronal}, running-difference \citep{attrill2010automatic} or direct \citep{krista2017statistical} images of the Sun in the EUV and SXR. These methods do not specifically target flux rope dimmings, and thus they often miss them. More specialised approaches aimed at identifying flux rope dimmings \citep{xing2020evolution, wang2023investigating} incorporate additional physical constraints, such as the use of flare ribbons, to delimit flux rope dimming regions. Still, these methods are often limited to relatively standard eruption scenarios in which observations follow established flare eruption models, i.e., they become hard to use or interpret in cases of complex eruptions. 

Characterising MFR footpoints is crucial for understanding the formation, evolution, and geo-effectiveness of solar eruptions. Flux rope dimmings, when they are observed on disc and where magnetic field measurements are available, provide key insights into the build-up and eruption of MFRs and the formation of CMEs. For example, flux rope dimming studies have revealed increases in MFR twist during the eruption, indicating the gradual accumulation of magnetic energy \citep{wang2017buildup, wang2019evolution} and the decrease of electric current \citep{barczynski2020electric} due to the expansion of the MFR. Flux rope dimmings have also been used to derive the magnetic flux contained in erupting MFR \citep{attrill2006using, cheng2016nature, Temmer2017, Wagner2023}, with the derived values being consistent with the flux measured in situ \citep{webb2000relationship, qiu2007magnetic}, even though flux erosion is expected as magnetic reconnection occurs along the CME propagation path (see e.g. \citet{pal2021uncovering}). During the eruptions, the drifting and deformation of flux rope dimmings can also trace magnetic reconnection processes and the evolution of toroidal flux \citep{xing2020evolution, Gou2023complete}, as well as the expansion of the associated CME \citep{webb2000relationship, hu2014structures, qiu2017gradual}. 

\begin{figure*}[t!]
     \centering
     \includegraphics[width=0.9\linewidth]{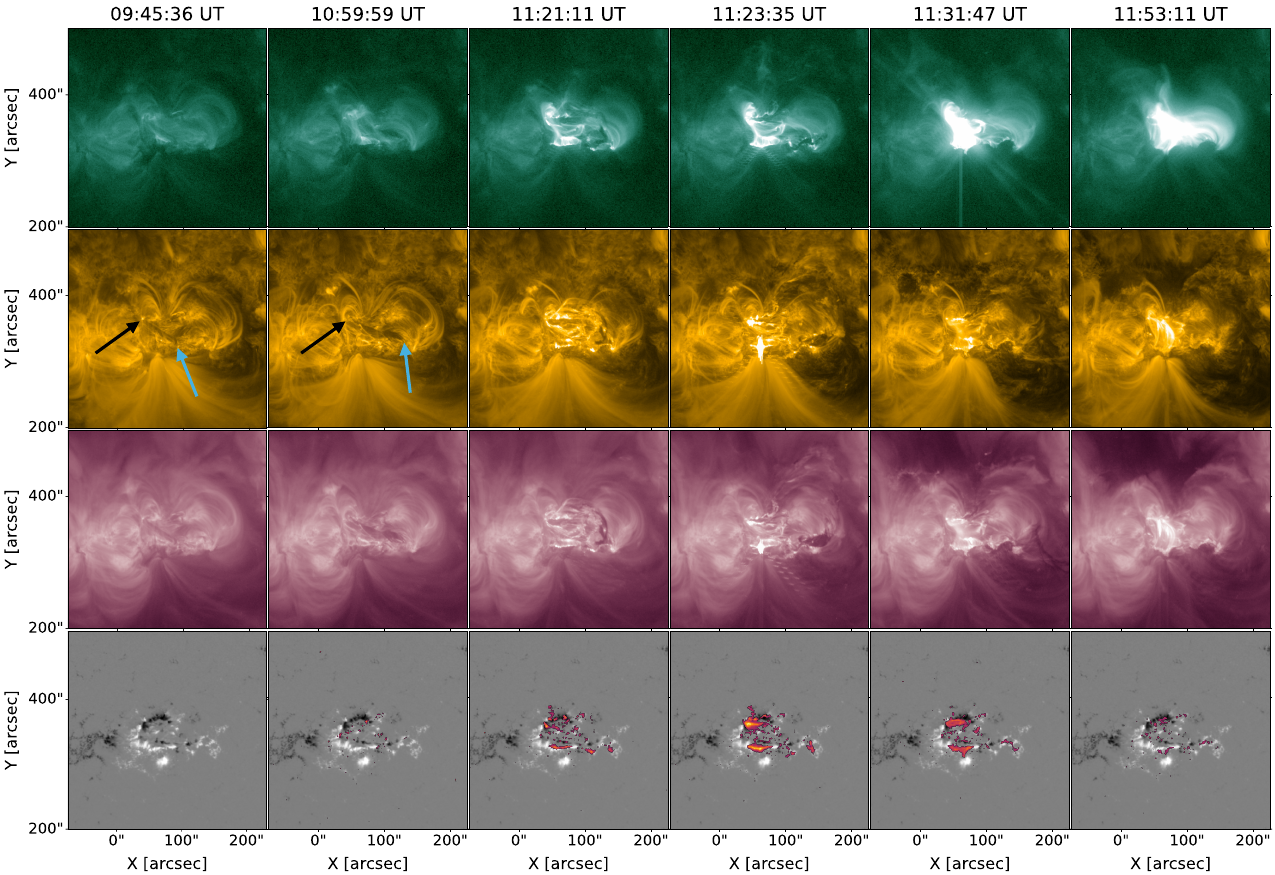}
     \caption{Overview of the M4 flare on 28 March 2022 as observed by SDO/AIA at six selected times (columns). From top to bottom, rows show AIA 94, 171, and 211~\AA~observations; bottom row shows AIA 1600~\AA~emission overplotted over an SDO/HMI radial magnetogram, where only emission above 100~DN~s$^{-1}$ is shown. The first column shows observations at 09:45~UT, before the preceding C-class flare. The black and blue arrows point to the filament footpoints. The associated movie is available online.}
     \label{fig:event_overview}
\end{figure*}

Relatively few studies have directly compared coronal dimmings with simulated MFR eruptions. Idealised models by \citet{gibson2008partially} and \citet{aulanier2019drifting} showed that the footpoints of erupting MFRs, and consequently the associated flux rope dimming signatures, can be effectively displaced by magnetic reconnection and that this motion follows quasi-separatrix layers \citep[QSLs;][]{Titov2002}.
Data-constrained simulations have further shown that coronal magnetic field topology could lead to remote flux rope footpoints outside of the active region \citep[AR;][]{lugaz2011numerical} or complex footpoint morphologies \citep{prasad2020mhd}. Simulations of the flare and CME event from AR~11158 on 15 February 2011 also showed that the strongest, long-lived dimmings were consistent with plasma evacuation in the MFR footpoints \citep{jin2022coronal} and that the flux rope dimming regions were located at regions where MFR field-lines were strongly stretched \citep{Fan2024}. This idea in particular was expanded on in \citet{kazachenko2026} where the authors used so-called L-maps -- i.e. maps of the natural logarithm of magnetic field-line lengths, to identify dimming regions in places where field-lines lengthen sufficiently strongly. 

To gain new insights into the complex nature of solar eruptions and related flux rope dimmings, we investigated the M4 flare and CME event that occurred in AR~12975 on 28 March 2022 (studied in detail in \cite{purkhart2023multipoint,Purkhart2024} to achieve three main objectives: (1) to perform a data-driven magnetofrictional simulation of the eruption and verify the model's magnetic field, and in particular, MFR evolution; (2) to examine how the detected flux rope dimmings correspond to the simulated MFR; and (3) to assess the limitations and caveats involved in identifying and interpreting flux rope dimmings in observations.

\section{Event overview and dimming analysis}
\label{sect:obs}

\subsection{Data and data reduction}
\label{sect:obs:data}
We used full-cadence data from the Solar Dynamics Observatory (SDO; \citealt{pesnell2012sdo})/Atmospheric Imaging Assembly (AIA; \citealt{Lemen2012aia}) to detect coronal dimmings and compared to the simulation. We detected the dimmings in the six AIA EUV channels (94, 131, 171, 193, 211, and 335~\AA), and in the 304~\AA~channel, which is sensitive to chromospheric temperatures. We used AIA 1600~\AA~data to analyse the spatial relationship of the flare ribbons, the flux rope footpoints, and the coronal dimmings. We only considered AIA images with exposures between 1.8~and~3.0~s. We derived the magnetic properties of the coronal dimmings with a single HMI vector magnetogram (\texttt{hmi.B\_720s}) taken at 10:45~UT. We computed the radial magnetic field, $B_r$, from the vector magnetogram using the IDL procedure \texttt{hmi\_b2ptr} following the method described by \citet{sun2013coordinate}.

We rebinned the images to $2048\times2048$ pixels under flux-conservation conditions, resulting in an effective spatial resolution of 1.2~arcsecs, and processed them with standard Solarsoft IDL software (\texttt{aia\_prep.pro} and \texttt{hmi\_prep.pro}). To co-register AIA and HMI data for the magnetic flux calculations of the dimming region, we used \texttt{coreg\_map.pro}. We differentially rotated all images to 11:00~UT. Since we focused the study on the flux rope dimmings, we studied coronal dimmings within a $400\times 400''$ sub-field around the AR.

\subsection{Event overview}\label{sect:obs:event}
The event under study is an eruptive M4 class flare that originated from AR~12975 on 28 March 2022. The flare started around 11:00~UT and reached its peak at 11:29~UT, and it was associated with a filament eruption and an Earth-directed CME \citep{purkhart2023multipoint}. Figure~\ref{fig:event_overview} and the accompanying movie shows the evolution of the event in the AIA 94, 171, and 211~\AA~wavelength channels. The bottom row shows an HMI radial magnetogram obtained 15~minutes before the eruption, overlaid with AIA 1600~\AA~emission at the corresponding times, where only intensities above 100~DN~s$^{-1}$ are displayed. 

At the beginning of the event, a highly sheared filament channel is visible in the second column of Fig.~\ref{fig:event_overview}. The filament gradually rises until it destabilises, and at 11:21~UT the northern footpoint detaches from the photosphere, which coincides with the development of flare ribbons on both sides of the central polarity inversion line (PIL). The southern footpoint remains anchored for several additional minutes before finally detaching around 11:34~UT. The flare ribbons initially develop in the two main strong polarities of the AR, but the southern footpoint is rooted in a weaker positive polarity region farther west, where additional flare ribbons can also be observed from 11:21~UT.

The erupting filament itself was formed during precursor activity associated with a confined C2 flare that occurred 1.5~hours earlier. \citet{Purkhart2024} showed that, prior to this small confined flare, two neighbouring filament channels were present in the AR. The first column of Fig.~\ref{fig:event_overview} shows this configuration, with the black and blue arrows pointing at the footpoints of the most prominent filament before the C2 flare. During the C2 flare, tether-cutting reconnection restructured these channels, reconnecting the southern footpoint of the main filament into a longer filament where the southern footpoint was displaced westward (see \citet{Purkhart2024}). The resulting configuration corresponds to the filament visible in the second column of Fig.~\ref{fig:event_overview}, where the blue arrow marks the newly formed southern footpoint of the filament. 

The event proved well-suited for our study for two major reasons: (1) it is very well-observed with clear observational signatures that can be directly compared with the simulation results, allowing us to robustly validate the model's accuracy to capture the magnetic topology; and (2) the important dynamics of the AR occurred sufficiently close to the disc centre. In particular, significant flux emergence in the form of a bipole took place around 10~UT on 26 March 2022, with the negative polarity region located to the east. Flux emergence continued over the following days, while the region migrated closer to the original positive polarity part of the AR. This negative polarity spot ultimately became the anchoring location for the northern footpoint of the filament and flux rope (see black arrows in the first two panels of AIA 171 \AA~(second row) and the magnetic configuration in the bottom row of Fig.~\ref{fig:event_overview}). Being able to follow this evolution and incorporate it in the data-driven modelling greatly enhanced our ability to accurately simulate the event.

\subsection{Coronal dimming detection} 
\label{sect:obs:dimming}

\begin{figure*}
     \centering
     \includegraphics[width=0.9\linewidth]{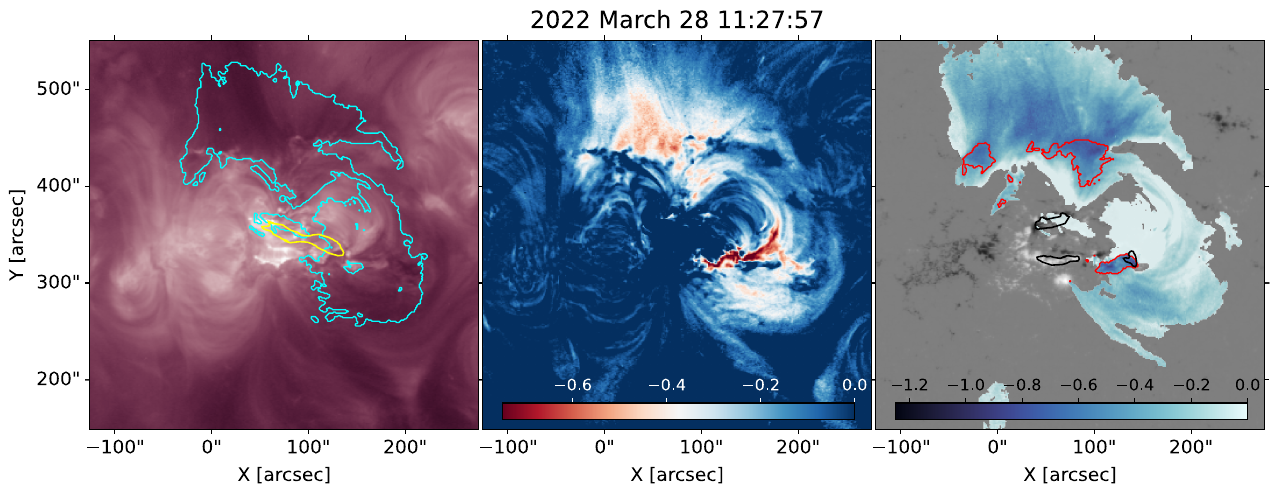}
     \caption{Evolution of the coronal dimming associated with the M4 flare on 28 March 2022. Left panel: SDO/AIA 211~\AA~direct image at 11:27~UT, with the dimming region detected up to that point contoured in cyan. Centre panel: Corresponding logarithmic base-ratio image. Right panel: Minimum-intensity map from logarithmic base-ratio data overlaid on an SDO/HMI radial magnetogram, where the red outlines indicate core dimming regions detected with the method of \citet{Dissauer2018a}. The yellow contour in the left panel shows the filament at 11:10~UT as observed in AIA 171~\AA, and the black contours in the right panel shows the main flare ribbon regions at 11:23~UT observed in AIA 1600~\AA~(see Fig.~\ref{fig:event_overview}). The associated movie is available online.} 
     \label{fig:overview_dimming}
\end{figure*}

We extracted coronal dimmings using the method developed by \citet{Dissauer2018a}, where a thresholding technique is applied to SDO/AIA 211~\AA~images. In this method, a pixel is considered a dimming pixel if its logarithmic base-ratio (log$_\text{10}$) intensity decreases below $-0.19$. We constructed the logarithmic base-ratio images using the average of three consecutive images at 10:45~UT as a base image. Pixels that fulfil the condition are cumulatively added to a dimming pixel mask. Figure~\ref{fig:overview_dimming} and the corresponding online movie show the overview of the coronal dimming detection for the event under study. The left panel shows the dimming pixel mask in cyan contours on top of the 211~\AA~image at 11:27:57~UT, which corresponds to the moment when the filament is almost completely detached and both flux rope dimmings are visible simultaneously, before the northern flux rope dimming closes. The middle panel shows the corresponding logarithmic base-ratio image at the same time step. The white and red regions mark areas where the brightness has strongly decreased; that is, the dimming regions. We create minimum-intensity maps that show the minimum intensity over the detection time of the pixels within the dimming mask \citep{thompson2016persistence}. Before the eruption, the filament is visible in the yellow-outlined region in the left panel of Fig.~\ref{fig:overview_dimming}, after which it whips westwards. To avoid pollution from the dark filament structure, we start calculating the minimum intensity map after the filament has fully erupted at 11:40~UT. The right panel of Fig.~\ref{fig:overview_dimming} shows the minimum intensity map in logarithmic base-ratio units computed on AIA 211~\AA~observations, where darker pixels are linked to larger mass depletions. We calculated the minimum intensity maps in base-difference and logarithmic base-ratio units. The temporal evolution of this figure is available as an online movie.

To identify flux rope dimmings, \citet{Dissauer2018a} developed a method using minimum-intensity maps, whereby a dimming pixel is considered to be a flux rope dimming if it is detected within the early impulsive phase of the total dimming (within 30~minutes) and its pixel intensity decreases below thresholds $A$ and $B$, calculated from base-difference and logarithmic base-ratio minimum-intensity maps, respectively. The authors defined the thresholds as follows:
\begin{equation}
    A=\bar{I}_\text{BD}-0.6\sigma_\text{BD} \\ \text{and} \\
    B=\bar{I}_\text{LBR}-0.6\sigma_\text{LBR},
\end{equation}
where $\bar{I}_\text{BD}$ is the mean intensity of the base-difference minimum intensity map, $\bar{I}_\text{LBR}$ is the mean intensity of the logarithmic base-ratio minimum-intensity map, $\sigma_\text{BD}$ and $\sigma_\text{LBR}$ are the corresponding standard deviations.

We applied the method of \citet{Dissauer2018a} to identify flux rope dimming regions. For the event under study, we did not use the temporal constraint because the filament remains visible during the initial phase of the dimming expansion. Consequently, we performed the flux rope dimming detection solely with intensity thresholds. The resulting flux rope dimming region as detected in AIA 211~\AA~observations is shown by red contours in the right panel of Fig.~\ref{fig:overview_dimming}. Figure~\ref{fig:overview_dimming_all} and its accompanying movie show the equivalent to Fig.~\ref{fig:overview_dimming} in all remaining wavelengths where we performed the detection (94, 131, 171, 193, 335, 304~\AA). This method yielded a single cumulative dimming mask per wavelength.

In the case of the northern footpoint of the filament, the method of \citet{Dissauer2018a} fails to identify a dimming region. This occurs because the filament lies on top of the region in the pre-event phase, and thus the intensity does not noticeably decrease (see the yellow contours in Fig.~\ref{fig:overview_dimming}). Furthermore, this footpoint is closely surrounded by flare ribbons, which quickly end up sweeping over the footpoint and closing the region; this renders methods relying on intensity minimums and cumulation of pixels not useful. The black contours in the right panel of Fig.~\ref{fig:overview_dimming} show the main flare ribbons at 11:23~UT, which are co-spatial with the northeastern footpoint of the filament.
To overcome this limitation, we implemented an alternative detection approach inspired by the methods from \citet{xing2020evolution} and \citet{wang2023investigating}. In these approaches, flare ribbons are used to constrain the spatial extent of flux rope dimmings, while localised brightenings serve as boundaries or source regions for the flux rope dimmings. 

\begin{figure}
     \centering
     \includegraphics[width=0.9\linewidth]{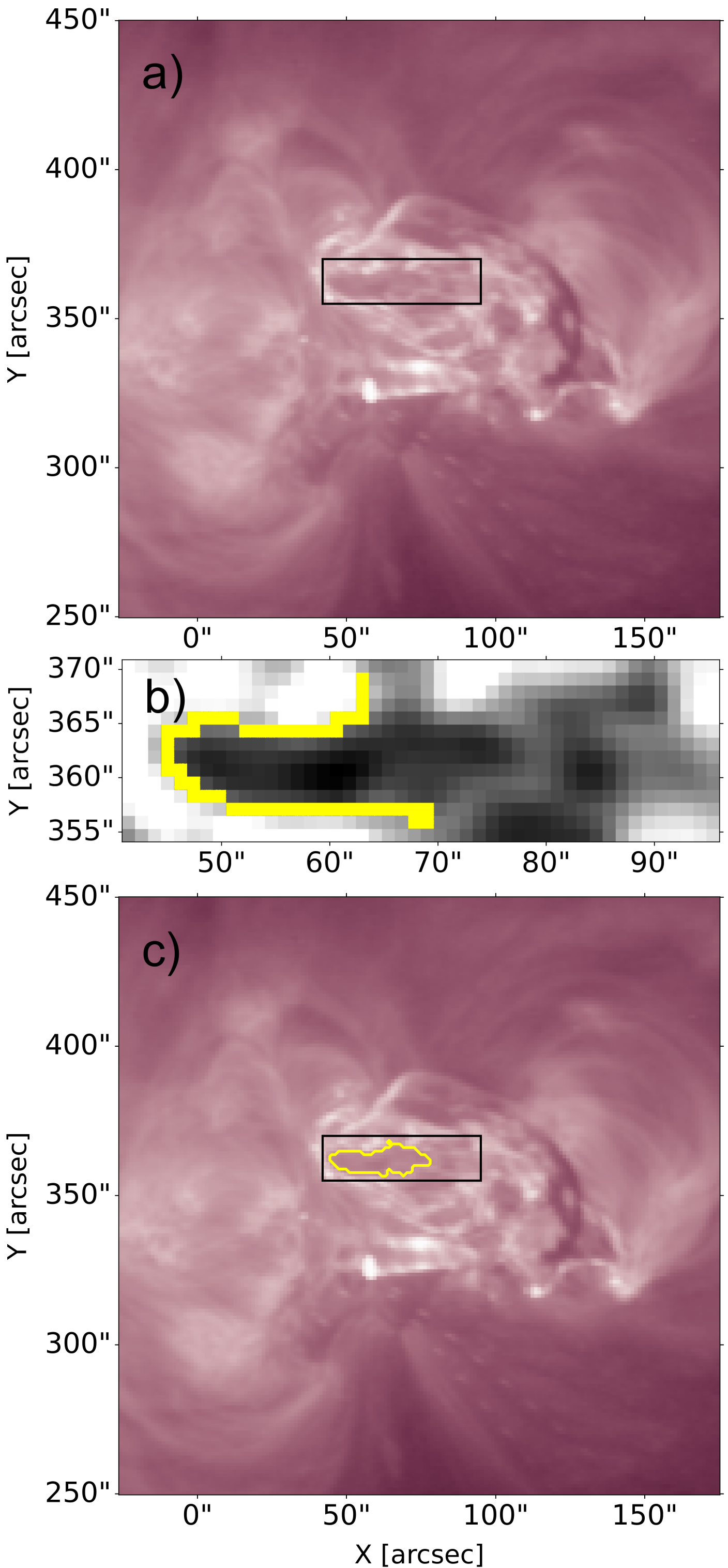}
     \caption{Method for the detection of the northeastern flux rope dimming exemplified with the AIA 211~\AA~channel. Panel a: AIA 211~\AA~image at 11:22:21~UT with a box encompassing the flux rope footpoint region where detection occurs. Panel b: Box region but with the intensity smoothed and cut off. The yellow line shows the strongest gradient region. Panel c: Same as top panel, but with a yellow line showing the detected footpoint region using a region growth algorithm from the line in the middle panel.}
     \label{fig:north_dimming}
\end{figure}

Specifically, for the northern footpoint, the detection procedure consists of the steps represented in Fig.~\ref{fig:north_dimming} for the observations in AIA 211~\AA. First, we selected a sub-region where the northern footpoint of the filament is completely captured (shown as a black square in Figs.~\ref{fig:north_dimming}a and c. Second, we applied a gaussian smoothing within the selected box and cut off the intensities above one standard deviation over the mean intensity (Fig.~\ref{fig:north_dimming}b). Third, we performed an edge search using a standard Canny detector within the scikit-image\footnote{\url{https://scikit-image.org}} Python package and selected the edge with the strongest gradient (yellow line in Fig.~\ref{fig:north_dimming}b). Fourth, we used the edge pixels as seed regions to grow the footpoint region, which is only allowed to grow towards the lower intensity side of the gradient and with a maximum intensity growth of 2\%. Figure~\ref{fig:north_dimming}c shows the footpoint region grown from the seed edge illustrated in panel b in yellow. This procedure yields a new flux rope dimming mask per time step for each wavelength. 

\begin{figure}
     \centering
     \includegraphics[width=\linewidth]{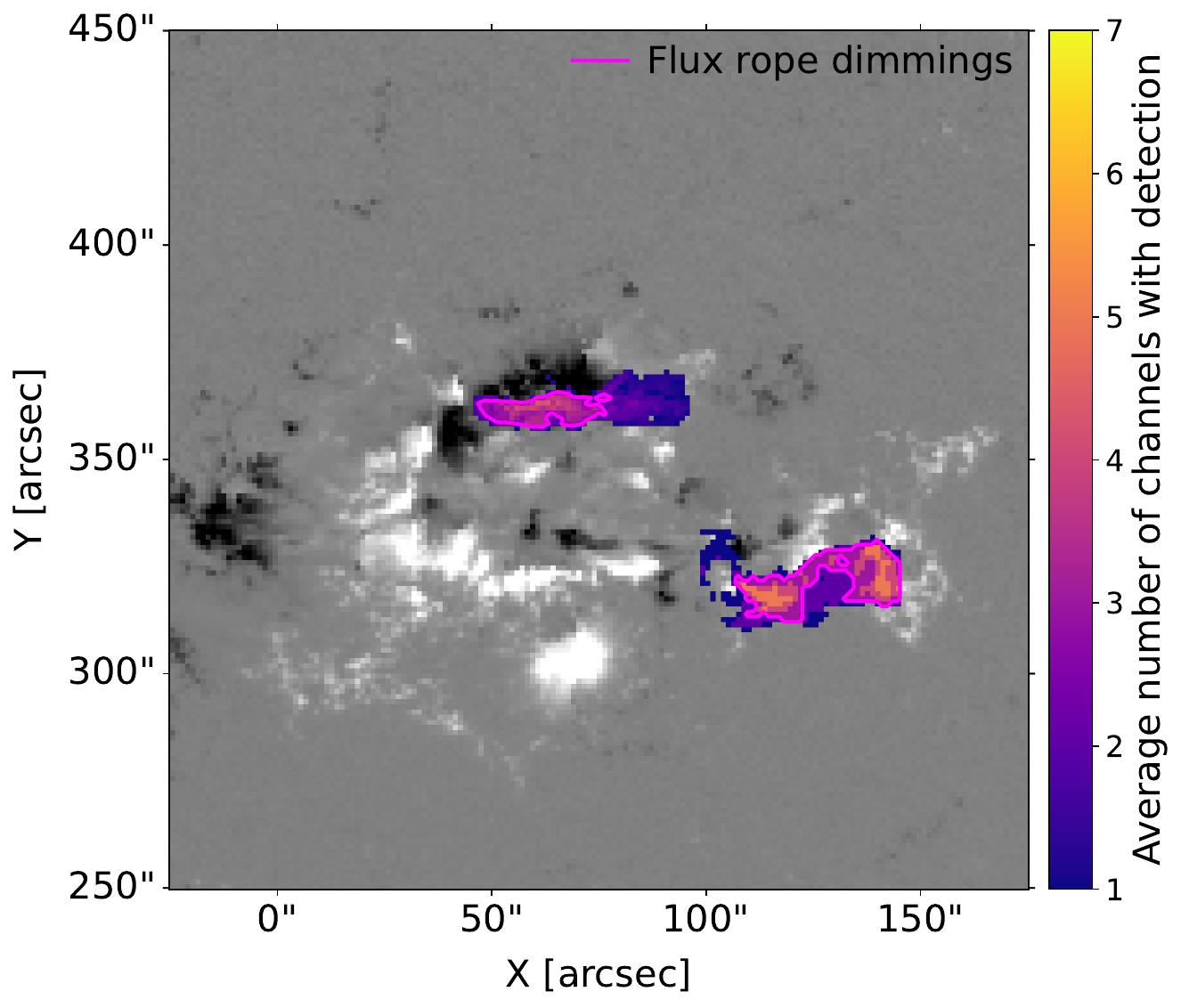}
     \caption{Detected flux rope dimming regions. The colour of each pixel indicates the average amount of AIA channels with detection during the detection period. A magenta line outlines the regions where three or more channels detect a dimming region. The associated movie is available online.}
     \label{fig:core_dimmings}
\end{figure}

Fig.~\ref{fig:core_dimmings} presents the cumulative masks of both flux rope dimmings extracted using the two independent detection methods. The mask colour indicates the number of AIA channels that identified a given region as a flux rope dimming. For the northeastern footpoint, the movie accompanying Fig.~\ref{fig:core_dimmings} shows the temporal evolution of the combined multi-channel detections, where the sweeping of the flare ribbons across the dimming region is clearly captured. As the footpoint subsequently closes, the algorithm begins to identify unrelated regions; therefore, we limited the valid detection time to be within 11:20 and 11:23~UT. We constructed a cumulative mask with the remaining time steps across all channels, which results in the northeastern footpoint region shown in Fig.~\ref{fig:core_dimmings}. To ensure that the selected regions correspond to plasma depletions, we defined the final flux rope dimming regions for both footpoints by requiring that at least three AIA channels detect a flux rope footpoint. These regions are outlined by the magenta contours in Fig.~\ref{fig:core_dimmings}.

\section{Modelling}
\label{Sect: model}
\subsection{TMFM simulation}
To obtain a more detailed picture of the magnetic topology of the AR, we used the time-dependent data-driven magnetofrictional model to simulate the magnetic field evolution of the AR \citep[TMFM;][]{Pomoell19}. In particular, we modelled the build up of free energy in the system, involving the self-consistent formation and destabilisation of an MFR. The magnetofrictional method builds on the prescription that the velocity in the modelling domain is proportional to the Lorentz-force \citep{Yang1986}: 

\begin{equation}
\mathbf{v} = \frac{1}{\nu}\frac{\mu_0 \mathbf{J} \times \mathbf{B}}{B^2}  .  
\end{equation}

The simulation was initialised with a potential magnetic field (starting time: 25 March 2022 at 17 UT) while its subsequent evolution is followed the magnetofrictional method. The starting time was chosen to capture as much of the evolution of the AR as possible, while avoiding significant degradation of the quality of the vector magnetograms due to the AR approaching the limb. Most importantly, this initiation time  also captures the full evolution of the AR from the onset of flux emergence until the eruption, as described in Sect.~\ref{sect:obs:event}. The model's temporal evolution is driven by photospheric electrograms, which are derived from magnetograms in the following way: First, photospheric velocity maps are derived from the temporal evolution of the vector magnetograms from SDO/HMI using the differential affine velocity estimator for vector magnetograms \citep[DAVE4VM;][]{Schuck2008}. From these maps, photospheric electrograms are computed using \[ \mathbf{E}_D = - \mathbf{v}_D \times \mathbf{B}, \] where the subscript `D' denotes that the quantity is derived with DAVE4VM. The resulting electric field maps are not directly used as input to the simulation as they are not necessarily consistent with Faraday's law. Instead, they serve as a reference for constraining the electrograms that are created by decomposing the electric field into an inductive and a non-inductive component. The inductive component can be derived from 
\begin{equation}
    \frac{\partial \mathbf{B}}{\partial t} = -\nabla \times \mathbf{E}_{ind}. 
\end{equation}
The non-inductive component ($\mathbf{E}_{nind} = \nabla \psi$) is then prescribed to follow the ad hoc assumption \citep[][]{Lumme2017}: 
\begin{equation}
    -\nabla^2_h\psi = U(\nabla \times \mathbf{B})_z.
\end{equation}
Here, $U$ is a free parameter, which is generally chosen such that the energy injection from \[\mathbf{E} = \mathbf{E}_{ind} + \mathbf{E}_{nind}\] matches the energy injection of $\mathbf{E}_D$ as
closely as possible. However, this particular setup assumes that all free energy of the AR is built up within the time span of the simulation and is spatially uniform. In our case, a better match with observations was achieved by a) including the neighbouring AR in the simulation domain, as there is a notable magnetic connection between the two and b) increasing the energy-optimised value of $U$ to $U' = 1.5U$ to facilitate a more realistic evolution of the AR's magnetic field. We then analysed the model results on a three-hour cadence. While this is a rather low cadence, particularly for eruption dynamics, we note that the dynamics in magnetofriction do not follow observed timescales in the eruptive phase of the MFR evolution. 
\begin{figure*}
     \centering
     \includegraphics[width=0.8\linewidth]{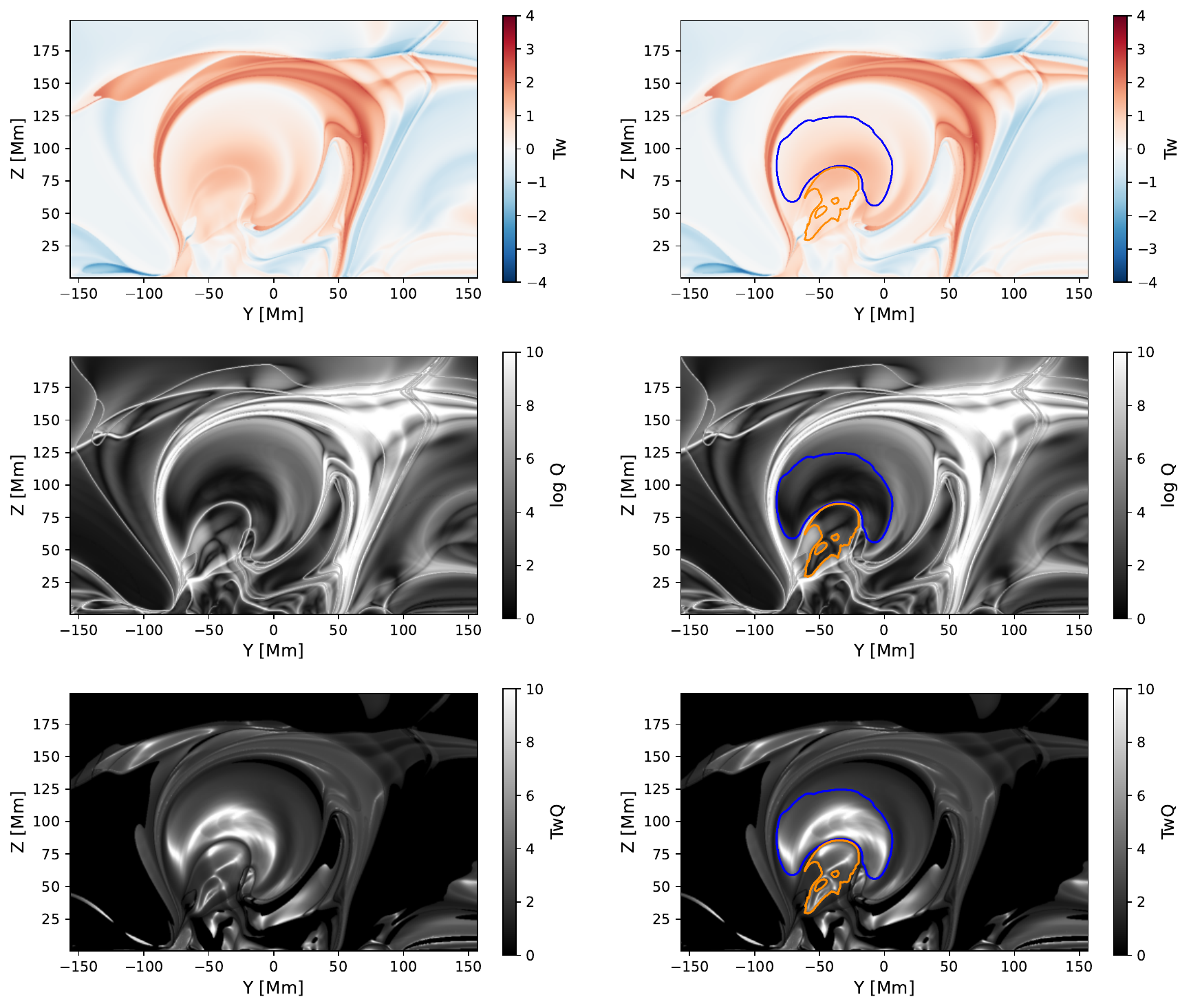}
     \caption{Showcase of the $T_w$ map, $log(Q)$ map and combined $T_Q$ map together with the extracted MFR structures of Fig.~\ref{fig: LateEvo} for Frame 30 (panels c) and d) of Fig.~\ref{fig: LateEvo}). The contours match the field-line colouration of Fig.~\ref{fig: LateEvo}.}
     \label{fig: MFRExtraction}
\end{figure*}
Thus, the evolution of the magnetic topology is followed qualitatively, independently of the exact timing of model snapshots in the later stages. We therefore did not refer to the timing of the derived boundary conditions but instead used frame numbers.

\subsection{Magnetic flux rope extraction}
\label{Sect:Extraction}
To extract the MFR field lines, we used the Graphical User Interface for Tracking and Analysing flux ropes \citep[GUITAR;][]{Wagner2024b}. The MFR extraction scheme implemented in GUITAR is based on the method described in \cite{Wagner2024a} and can be summarised as follows: First, a suitable plane is selected within the simulation domain at a location where the MFR cross-section is expected to be fully contained. Next, an MFR proxy of choice is computed on this slice, after which a thresholding procedure is applied  to obtain the first estimate of the MFR cross-section. The resulting contours are then post-processed with mathematical morphology (MM) operations to more accurately capture the actual MFR cross-section. In particular, we performed a series of opening algorithms to remove noisy features at the cross-section edges, as well as any  unwanted components connected to the MFR cross-section. In contrast to previous studies, we used a combination of the twist parameter $T_w$ and the logarithm of the squashing factor $\log(Q)$ to obtain improved results, particularly for disentangling flux systems in the later stages of the simulation. $T_w$ quantifies the turns a field-line takes around an infinitesimally close neighbouring field-line and can be computed as follows:
\begin{equation}
  T_w = \int_{L} \frac{\mu_0 J_{\parallel}}{4 \pi B} \, dl.
\end{equation}
It generally attains high values within the MFR cross-section, as MFRs are defined as coherent bundles of twisted field lines. High $Q$ values on the other hand indicate a strong change in magnetic connectivity and regions of high $Q$ are associated to QSLs \citep{Titov2002}. $Q$ is a particularly relevant quantity in the context of MFRs as they are expected to be coherent structures of similar magnetic connectivity, while  the surrounding fields typically exhibit notably different connectivity. Therefore, QSLs are commonly used in identifying MFRs in magnetic field simulations \citep[see e.g.,][]{PJZhang2022}. One downside of this parameter is that for very dynamic and complex magnetic field simulations, maps of $\log(Q)$ may appear highly structured and extracting the desired information may be non-trivial. However, the additional information gain through sharper outlines provided by $\log(Q)$ proved to be a helpful addition for extracting the flux systems of the TMFM simulation of AR~12975. Thus, an empirical combination of $T_w$ and $\log(Q)$ was constructed in the following way: 
\begin{equation}
    T_Q \equiv M \left[ T_w + \frac{c}{\log(Q)} \right],
\end{equation}
where $M$ is a binary mask that removes all regions with $T_w$ polarities opposite to the (suspected) MFR twist polarity. This has the advantage of removing some of the complexity introduced by the inclusion of $\log(Q)$. $c$ is a constant that appropriately scales the $\log(Q)$ values to $T_w$ (here, $c = 6$). We find that this construction contains both of the large-scale features from $T_w$, but it also preserves the small-scale features and sharp edges from the $\log(Q)$ maps.

A comparison of $T_w$, $\log(Q)$ and $T_Q$ maps is provided in Fig.~\ref{fig: MFRExtraction} along with the extracted contours in the right-hand panels for one example snapshot. The $T_w$ maps highlight the large-scale high-twist features present in the simulation, but especially in the regions of interest, edges are not particularly sharp and thus hard to identify. On the other hand, the $\log(Q)$ maps depict many of these edges rather well; however, it does not contain any information on the field-line twist or twist polarity. In contrast, the combined metric highlights the regions of interest remarkably well (bottom panels of Fig.~\ref{fig: MFRExtraction}), and serves as the final result from which the MFR cross-section contours are derived. Once the cross-section masks have been created, points are sampled uniformly within them to calculate the MFR field lines. Lastly, to derive the footpoints of the MFR, we computed the intersection of its field-lines with the bottom boundary of the simulation domain.

\subsection{Characteristic properties}\label{sec:properties}
To compare the characteristics of the MFR between the TMFM simulation and the observations, we used an observational approach. For the observed flux rope dimmings we first extracted the total area ($A$) of the dimming masks. We then calculated the magnetic area ($A_\phi$), which is the flux rope dimming region where the radial magnetic field $B_r$ has absolute values above the noise level, i.e., $|B_r|>100$~G, following \citet{Kazachenko2017database}. From the magnetic area we then calculated the unsigned magnetic flux ($\phi$) and the mean magnetic flux density ($B$) (see \citet{Dissauer2018a} for more information on the dimming parameters).  

Noting that in the magnetofrictional framework the eruption dynamics do not evolve on the observational timescales (see Sect.~\ref{Sect: model}), we estimated the relevant simulation time steps from the restructuring of the magnetic field. In particular, we combined the footpoint masks of the first three time steps after the MFR footpoints have fully migrated to the remote southern polarity region to construct the model flux rope dimming mask. 

By projecting the simulated flux rope footpoints onto the HMI radial magnetogram, we also derived the magnetic area, unsigned magnetic flux, and mean magnetic flux density within the simulated footpoint regions using the same observational analysis procedure. This way, we obtained a direct comparison to the dimming results. 

\section{Results}
\label{Sect: results}

\subsection{Evolution of the simulated flux rope} 

\begin{figure}
     \centering
     \includegraphics[width=\linewidth]{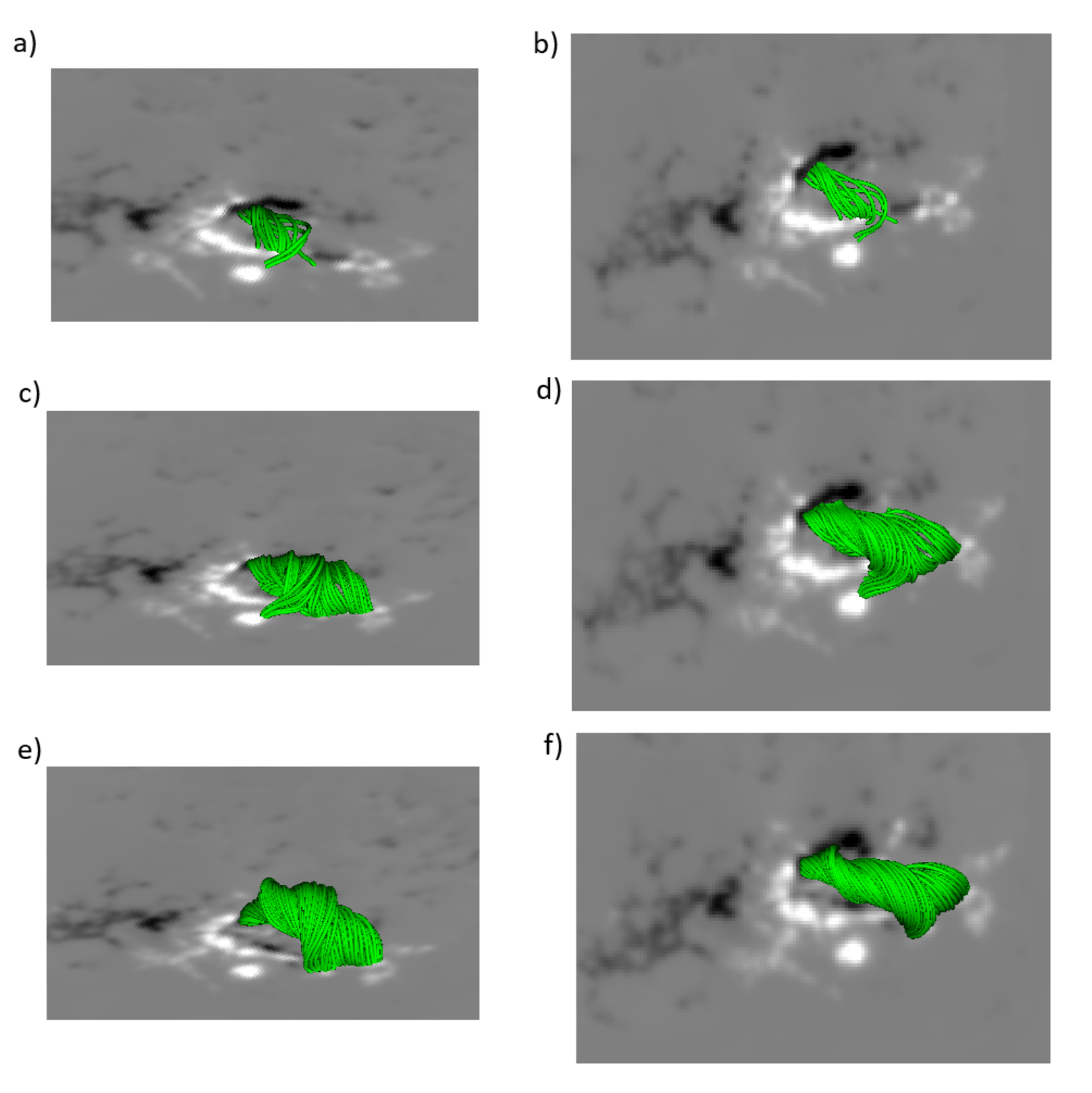}
     \caption{Early evolution of AR~12975 flux rope field lines. Panels a), c) and e) show the flux rope in simulation frames 19, 20 and 22, while panels b), d) and f) show the same field-lines (respectively) from a top view.}
     \label{fig: EarlyEvo}
\end{figure}

The magnetofrictional simulation captures the whole process from the MFR formation to its eruption from AR~12975. The early-stage evolution is shown in Fig.~\ref{fig: EarlyEvo}. In the very early stages, the magnetic field in the AR is close to potential, and thus it does not carry any significant twist. Consequently, the $\log(Q)$ maps show no clear indications of an MFR. After about two days of evolution a bundle of field-lines forms that accumulates twist as the simulation progresses. While GUITAR can detect early bundles of weakly twisted field lines, these features may not be most representative of the overall magnetic topology near the eruption site as clear MFR boundaries are not well identifiable at this stage in the proxy maps. Therefore, the field-lines depicted in the first row of Fig.~\ref{fig: EarlyEvo} were hand-picked. However, for the following times when GUITAR results are used (e.g., shown in the middle and lower panels of Fig.~\ref{fig: EarlyEvo}), the empirical metric introduced in Sect.~\ref{Sect:Extraction} is utilised. In the very early stage, the pre-eruption filament channel is outlined well by the field-lines shown in the top row panels, connecting the observed filament footpoint regions (see first column of Fig.~\ref{fig:event_overview}). 

\begin{figure}
     \centering
     \includegraphics[width=0.95\linewidth]{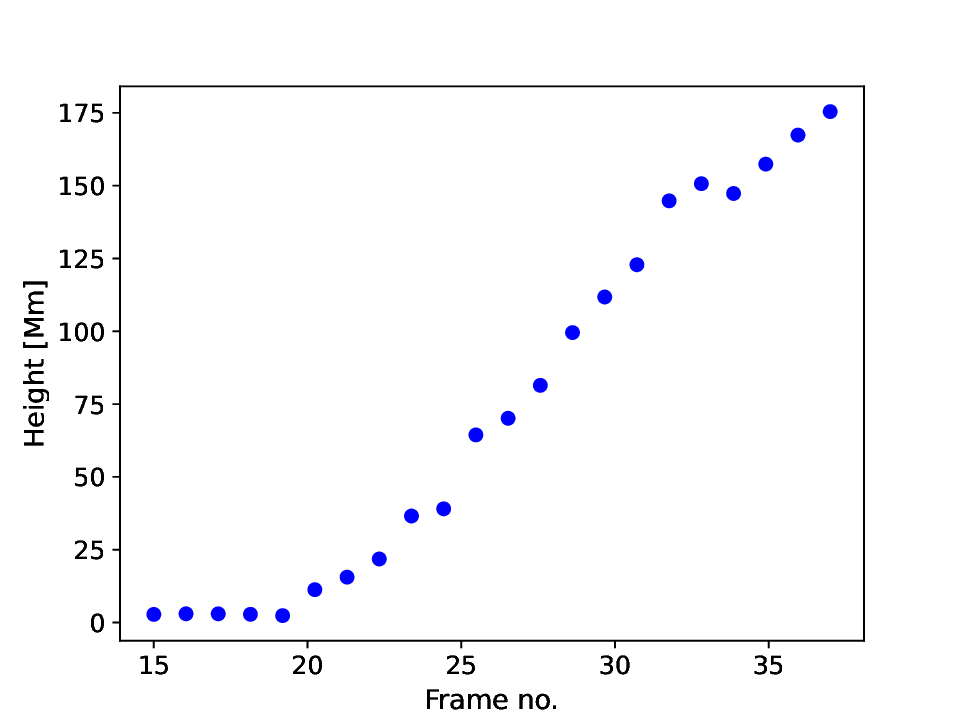}
     \caption{Evolution of MFR height, as averaged by all field-lines reaching above 10~Mm within the `true' apex.}
     \label{fig: FRHeight}
\end{figure}

Over the course of the simulation, a magnetic connection forms between the original filament footpoint in the north and the remote spot of positive polarity in the south that later becomes the new southern filament footpoint. As the bundle expands and the magnetic connection strengthens, it also becomes increasingly non-potential, forming an MFR in the process, as can be seen in the middle panels of Fig.~\ref{fig: EarlyEvo}. After the MFR footpoints in the south have fully migrated towards the western positive polarity spot, the MFR exhibits a visibly coherent structure (see bottom panels of Fig.~\ref{fig: EarlyEvo}). It is also during this restructuring that the large-scale dynamics of the MFR set in and the system appears notably unstable as the MFR begins to accelerate. To emphasise this, we calculated the evolution of the MFR height in Fig.~\ref{fig: FRHeight}. It is apparent that the MFR is accelerating around frame 20, coinciding with the onset of footpoint reconfiguration (see panels c and d in Fig.~\ref{fig: EarlyEvo}). Note that the height values displayed represent an average of all field-lines that reach within 10~Mm of the MFR apex. This choice was made to track the rise of the MFR front's centre of mass instead of individual field-line apexes. 

\begin{figure}
     \centering
     \includegraphics[width=\linewidth]{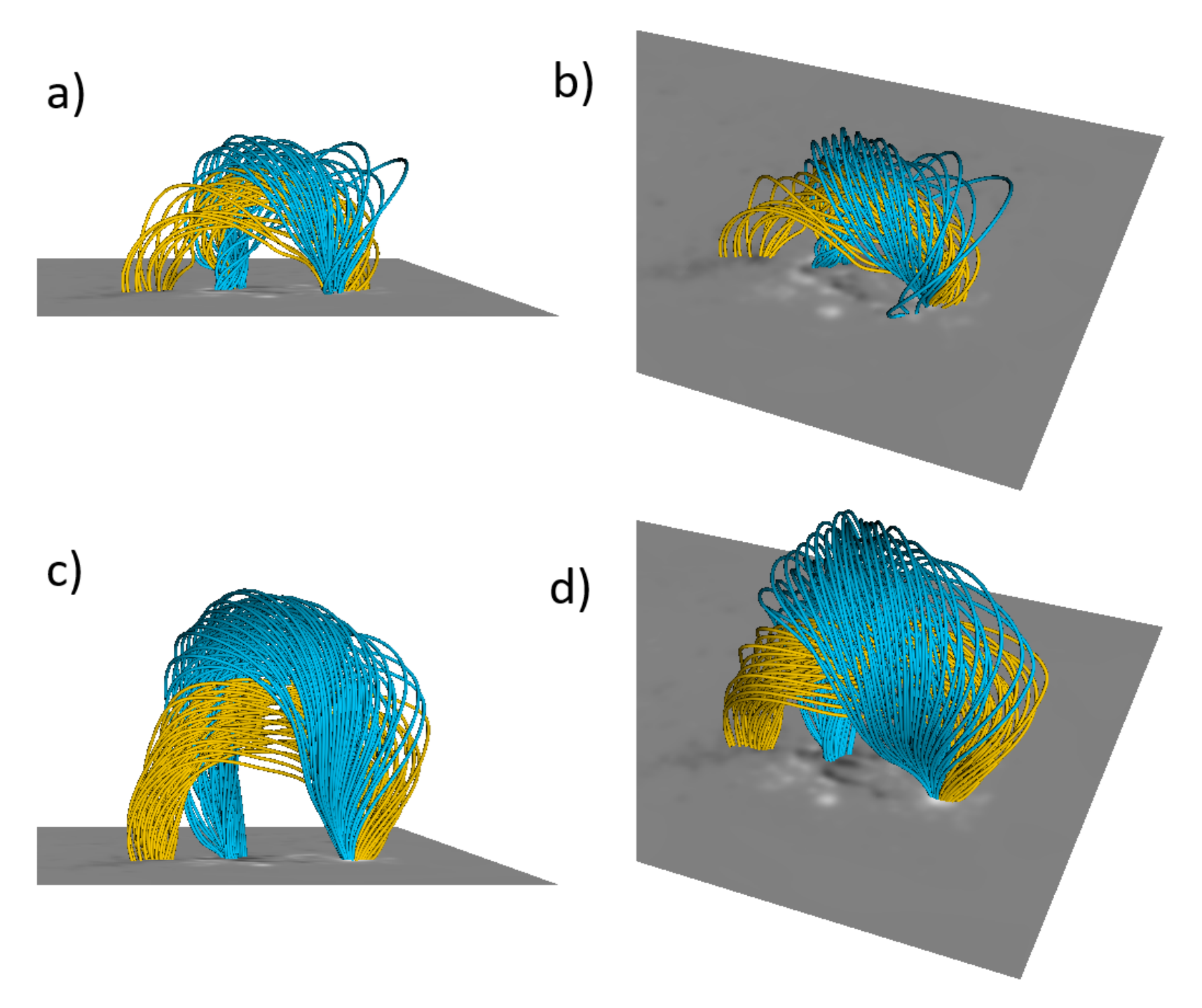}
     \caption{Mid- to late-stage evolution of AR~12975 field lines. Panels a) and b) depict field-lines identified at frame 27, while panels c) and d) show flux rope field-lines identified at frame 36.  The flux rope can be divided into two distinct sub-features, which are coloured with orange and blue field-lines (matching the outlines in Fig.~\ref{fig: MFRExtraction}). Left-hand and right-hand panels show different viewing angles. The field-lines shown were extracted with the method described in Sect.~\ref{Sect:Extraction}. }
     \label{fig: LateEvo}
\end{figure}

In the later stages, the height evolution is approximately linear, indicating that the simulated MFR rises rather steadily. From this point onwards, the MFR can be separated into two individual MFR systems as indicated in Fig.~\ref{fig: MFRExtraction} by the blue and orange contours. The corresponding magnetic field-lines are shown in Fig.~\ref{fig: LateEvo} with the same colours as the contours. Visually, the MFR maintains its large-scale structure while rising and expanding through the simulation domain in this later stage. 

\subsection{Observed flux rope dimmings}\label{sec:res:observations}

In Fig.~\ref{fig:core_dimmings}, the coloured regions mark the detected dimming areas, with the colour scale indicating the number of wavelength channels that identified a given pixel as part of the dimming region. Pixels detected in at least three channels are defined as the final flux-rope dimming region and are outlined by the magenta contour. 

The northern flux rope dimming can be classified as a shrinking flux rope dimming within the framework proposed by \citet{veronig2025coronal}. Such dimmings are characterised as stationary flux rope dimmings that are swept over by flare ribbons and may eventually be fully closed by them. Physically, this results from the reconnection of the erupting flux rope, for example through leg--leg reconnection, which progressively closes the magnetic flux rooted in the dimming region. In the present event, the flare ribbons quickly sweep across the northern flux rope dimming region until the dimming completely closes. NLFF extrapolations performed by \citet{purkhart2023multipoint} showed that the pre-event magnetic field-lines vault above the filament channel, connecting the east-west positive polarity band south of the AR to the northern negative polarity region. As the MFR expanded these overlying field-lines stretched and reconnected, producing the observed flare ribbons. \citet{purkhart2023multipoint} showed that the filament itself was most likely also involved in the reconnection process, which aligns with the closing of the northern flux rope dimming.

In contrast, the southern flux rope dimming exhibits characteristics of moving flux rope dimmings. These arise when the erupting flux rope reconnects with surrounding closed magnetic flux, causing its footpoint to migrate away from its original location, and resulting in an observable motion of the dimming. The southern flux rope dimming first appears below the dense filament material as the filament erupts, and it subsequently displays southward motion (see movie accompanying Fig.~\ref{fig:event_overview}). This evolution is accompanied by the formation of flare ribbons and newly reconnected loops in the northern portion of the original dimming region. At the same time, widespread dimmings can be observed expanding north from the AR (see Fig.~\ref{fig:overview_dimming}) alongside propagating brightenings (see Fig.~\ref{fig:overview_dimming_all}). These observations suggest that the MFR reconnects with exterior magnetic flux located north of the AR, which facilitates the complete closure of the northern flux rope dimming. 

\subsection{Appearance and characteristic properties of the flux rope dimming region and MFR footpoints}

\begin{figure}
     \centering
     \includegraphics[width=\linewidth]{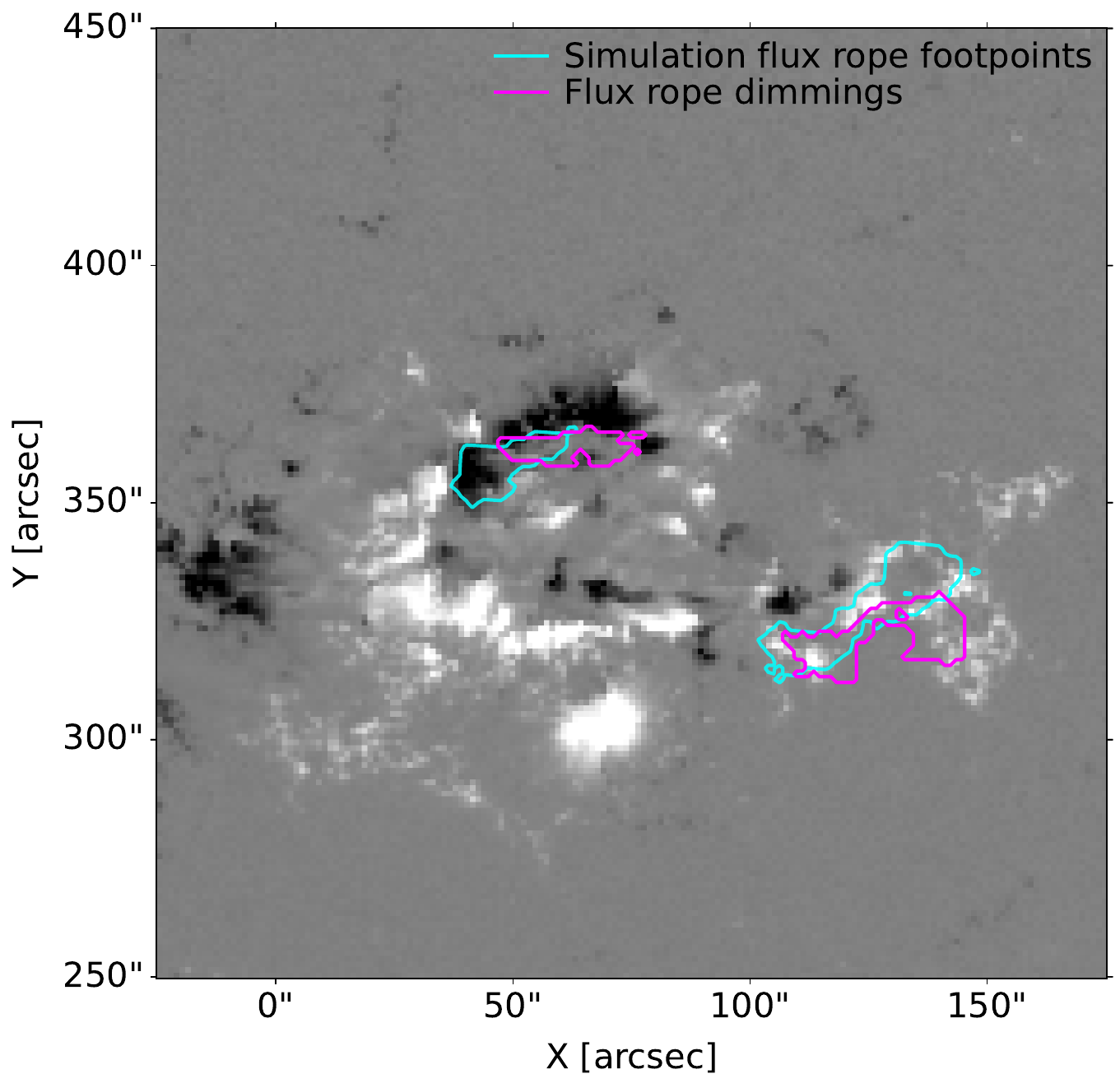}
     \caption{Comparison between the flux rope footpoint regions as extracted from the TMFM simulation (cyan) and the flux rope dimming observations determined from SDO/AIA data (magenta; see also Fig.~\ref{fig:core_dimmings}). The footpoints are shown as coloured contours over an HMI radial magnetogram.}
     \label{fig:core_dimmings_comp}
\end{figure}

We compare both of the derived core or flux rope dimmings from observations with the footpoint regions extracted from the TMFM simulation. As described in Sect.~\ref{sect:obs:dimming}, the dimming regions were selected as the cumulated detection mask where at least three AIA channels consider a pixel as part of the dimming region; for the simulation footpoints we combined the masks of the three time steps after the MFR had fully migrated to the remote southern polarity region (see Sect.~\ref{sec:properties}). Specifically, we used the time steps 21 to 23 to extract the simulated MFR footpoints; that is, the flux rope magnetic field-lines shown in Figs.~\ref{fig: EarlyEvo}e and f. The resulting footpoint regions are displayed in Fig.~\ref{fig:core_dimmings_comp}. The observed flux rope dimming areas agree generally well with the MFR footpoints in both the northeastern and southwestern regions. 

\begin{table*}
    \caption{\label{table:properties}Properties of the flux rope footpoints.}
    \centering
    \begin{tabular}{cccccc}         
    \hline\hline          
    Source & footpoint & \CellWithForceBreak{$A$ \\ ($10^{8} \text{km}^2$)} & \CellWithForceBreak{$A_\phi$ \\ ($10^{8} \text{km}^2$)} &  \CellWithForceBreak{$\phi$ \\ ($10^{20} \text{Mx}$)} & \CellWithForceBreak{$B$ \\ (G)} \\ 
    \hline
    \multirow{2}{*}{Coronal dimmings} & 
    North-east & $0.58$ & $0.29$ & $ 1.13 $ & $303$ \\ &  
    South-west &  $1.26$ & $0.66$ & $ 2.56 $ & $312$ \\
    
    \multirow{2}{*}{TMFM Simulation} & 
    North-east &$ 0.57 $&$ 0.50 $&$ 2.94 $&$ 515 $\\ &  
    South-west &$ 1.46 $&$ 0.71 $&$ 3.53 $&$ 244 $\\ 
    \hline 
    \end{tabular}
    \tablefoot{We list the total dimming area $A$, magnetic area $A_\phi$, unsigned magnetic flux $\phi$, and mean magnetic field density $B$ for the footpoints of the erupting flux rope as extracted from the observed flux rope dimmings and the TMFM simulation.}
\end{table*}

To further quantify similarities and differences between the observed flux rope dimmings and the simulated MFR, characteristic magnetic properties of these features were computed and compared. The results are summarised in Table~\ref{table:properties}, which shows the dimming area $A$, magnetic dimming area, $A_\phi$, unsigned magnetic flux, $\phi$, and mean magnetic flux density, $B$, of the separate flux rope dimming regions marked with magenta contours in Fig.~\ref{fig:core_dimmings}. The northern dimming region is considerably smaller in terms of both dimming area ($0.58\times10^8$~km$^2$) and magnetic dimming area ($0.29\times10^8$~km$^2$) than the southern dimming region, which is more than twice the size with values of $1.26\times10^8$~km$^2$ for the dimming area and $0.66\times10^8$~km$^2$ for the magnetic dimming area. The ratio between the two flux rope dimmings is similar for the unsigned magnetic flux, where the northern dimming has values of $1.13\times10^{20}$~Mx and the southern dimming has values of $2.56\times10^{20}$~Mx. These are at the lower end of the magnetic fluxes found by \citet{wang2023investigating} and within those found by \citet{Dissauer2018b}. The mean magnetic flux density, on the other hand, is very similar for both flux rope dimmings, i.e. 303~G and 312~G for the northern and southern dimmings, respectively. These values are comparable to magnetic field strengths in the vicinity of MFR footpoints in other simulations \citep[e.g.,][]{Sieyra2026}. 
 
\section{Discussion}
\label{Sect: discussion} 
\subsection{Comparison of observed flux rope dimmings to simulated flux rope footpoints}

The discrepancies between the simulated and observed footpoint locations in Fig.~\ref{fig:core_dimmings_comp} can be partly attributed to the evolution of the flux rope dimmings, which is not fully captured by the used detection methods. As discussed in Sect.~\ref{sec:res:observations}, the northern dimming is a shrinking flux rope dimming that is swept northwestwards by the flare ribbons. While this motion is captured by the instantaneous masks shown in the movie accompanying Fig.~\ref{fig:core_dimmings}, we used cumulative masks over several time steps to better constrain the dimming region, which excludes the initial eastern part of the dimming. Similarly, the southern flux rope dimming is identified from minimum intensity maps computed after the filament has fully erupted. By this stage, the southward motion of the moving flux rope dimming has already occurred, and the northern part of the dimming region has partially recovered. Consequently, the observed southwestern footpoint appears further south than its actual location. However, a comparison of the initial flare ribbons in Fig.~\ref{fig:event_overview} (third column) with the simulated MFR footpoints shows that the ribbons coincide with the eastern part of the northeastern footpoint location. Therefore, no dimming region could be detected there, regardless of the detection method used.

The simulated and observed MFR footpoints show very similar total areas, which are displayed in Table~\ref{table:properties}, particularly the northeastern footpoint, with $A=0.57\times10^8$~km$^{2}$ and $A=0.58\times10^8$~km$^{2}$, respectively. Differences become more apparent when comparing the magnetic areas. While the southwestern footpoints have similar magnetic areas ($A_\phi=0.71\times10^8$~km$^{2}$ simulated against $A_\phi=0.66\times10^8$~km$^{2}$ observed), the northeastern footpoint shows a more pronounced discrepancy, with $A_\phi=0.50\times10^8$~km$^{2}$ in the simulation and $A_\phi=0.29\times10^8$~km$^{2}$ in the observations. This occurs because the observed footpoint is located in a weaker magnetic field region, half of which is too weak to be considered within the magnetic dimming area, whereas the simulated footpoint is rooted in a stronger polarity region. As discussed earlier, this is a consequence of the initial easternmost part of the MFR footpoint not being captured observationally. Consequently, the unsigned magnetic flux differs most strongly for the northeastern footpoint, with the simulated footpoint ($\phi=2.94\times10^{20}$~Mx$^{2}$) containing more than twice the magnetic flux than the observed one ($\phi=1.13\times10^{20}$~Mx$^{2}$). It is also interesting to note that in the simulation, almost the whole extent of the compact northeastern footpoint region is located in strong field regions, while in the more spread out southwestern footpoint region the MFR has only about half of its field-lines rooted in strong field regions. The latter is very consistent with the observed ratio of total area to magnetic area, while for the northeastern region the discrepancy is yet again explicable by the shrinking nature of the flux rope dimming via the flare ribbons sweeping over the region. 

\subsection{Implications from the simulation}

The case of AR~12975 is particularly favourable for our modelling efforts. The AR and associated eruptive event exhibits well-defined, clearly observed features that allowed an accurate derivation of model parameters (e.g., energy injection, defining a suitable domain extent) and robust validation of our results. In particular the prominent filament and its restructuring provides strong constraints on the magnetic topology of the AR and its evolution. While the study demonstrates good agreement between the MFR footpoints inferred from the TMFM simulation and the observed dimming regions, it would be beneficial to reproduce the observed emission directly, for example using radiative MHD simulations \citep{Cheung2019}.

This approach would open the possibility to connect the flux rope dimming and the MFR within the model itself and could be a valuable addition to our understanding of how these dimming regions form. While the TMFM results show a remarkable match with the magnetic topology and its evolution of the AR as inferred from the observations, a more realistic dynamic evolution in the eruptive phase could also provide insight into the processes determining the overall timescales and different longevity of dimming regions within the event.

\subsection{flux rope dimming detection potential and limitations}

Despite the growing number of detection methods, identifying flux rope dimmings remains challenging because their observational signatures strongly depend on the eruptive configuration, line-of-sight effects, and the evolution of surrounding coronal structures.
Filaments are particularly problematic for methods based on intensity decrease, as they are intrinsically dark features in coronal wavelengths. Their presence creates low-intensity regions in pre-event images, which may not exhibit a sufficient decrease in intensity after the eruption. This was the case here for the northeastern footpoint of the studied filament, which was therefore not properly captured as a flux rope dimming region in most wavelength channels. Similarly, the motion of the filament can obscure dimming signatures, generating detections unrelated to plasma evacuation. One way to mitigate this issue, as implemented in this study, is to apply the thresholding procedure only after the filament has fully detached. The drawback of this approach is that it inevitably misses potential changes in the MFR footpoints. In fact, in the present event, by the time the filament was detached, the northeastern dimming had already closed and could not be detected.

An alternative strategy to avoid filament contamination is to identify only regions that persist sufficiently to be classified as dimmings. Since filament motion is typically fast, restricting the detection to pixels with sustained intensity decreases would help isolate stationary dimming regions. However, such a method would also fail to capture shrinking and moving flux rope dimmings. Likewise, simple intensity-thresholding approaches, such as the \citet{Dissauer2018a} approach employed here for the southwestern footpoint, can effectively identify the footpoint region, but it may also detect secondary dimmings whenever the intensity decrease is sufficiently strong. 

For example, the red contours in Fig.~\ref{fig:core_dimmings} showing the flux rope dimmings detected with the above discussed method, outline secondary dimming regions north of the AR. \citet{krista2017statistical} argued that dimmings could be better captured using direct images with methods similar to those used for coronal hole detection. However, this approach is rarely applicable to flux rope dimmings, which frequently occur within ARs with high brightness levels and complex structures. Other approaches, such as the persistence maps proposed by \citet{thompson2016persistence}, which show the lowest intensity of each pixel within the detection time, highly depend on the general evolution of the eruption. For example, in the present case, the filament would completely obscure any information since it sweeps over both footpoint regions.

For flux rope dimmings specifically, physically informed methods may provide a more reliable alternative. For example, \citet{xing2020evolution} and \citet{wang2023investigating} proposed approaches based on the assumption that eruptions evolve according to the standard flare model, whereby flux rope footpoints are expected to lie within J-shaped flare ribbons and coincide with regions of strong magnetic field. These methods are, however, only effective for the well-behaved eruptive events, as the event studied here illustrates that complex flare ribbons may not always directly point towards the MFR footpoints. The complex pre-eruptive magnetic configuration, shaped in part by the previous C-class flare, results in an asymmetric filament configuration, in which one MFR footpoint lies on the `expected' PIL side, while the other is displaced farther from the main PIL and located in a region of a comparatively weak magnetic field.

Overall, each detection method may be advantageous under specific observational settings. Identifying the limitations of a given method in a particular case requires a detailed understanding of the physical processes responsible for coronal dimming formation. In this regard, the classification framework proposed by \citet{veronig2025coronal} provides a useful scheme to facilitate the interpretation of coronal dimming observations alongside flare ribbon evolution.

\section{Conclusion}
\label{Sect: conclusion}
In this study, we investigated the evolution of the eruption-associated dimmings in AR~12975 and tested the common perception that core dimmings are outlining the erupting MFR's footpoints. 
The simulations capture the evolution of the observed features of the pre-eruptive structure in AR~12975 remarkably well: after about two days a bundle of field-lines forms that magnetically connects the observed filament footpoint regions. As the simulation progresses and an MFR forms from the bundle of weakly twisted loops, the footpoints of the MFR migrate fully towards the secondary footpoint region in the south and the system becomes seemingly unstable as a result of this magnetic reconfiguration. The observations show that the flux rope dimming regions undergo a complex evolution, with the northeastern region exhibiting a shrinking flux rope dimming that fully closes and the southwestern region behaving as a moving flux rope dimming. This evolution, alongside the obscuration caused by the erupting filament itself, influences the detectability and apparent morphology of the flux rope dimming regions. However, the derived footpoints of the simulation match the observed flux rope dimmings well, though discrepancies arise as the observed flux rope dimming in the northeast is swept over by the flare ribbons, possibly obscuring their full extent. 

Overall, this work highlights both the diagnostic potential and the observational limitations of flux rope dimmings, and emphasises the importance of combining observations, magnetic field modelling, and physically informed classification frameworks to correctly interpret dimming evolution in eruptive solar events.

\begin{acknowledgements}
This project has received funding from the European Union's Horizon Europe research and innovation programme under grant agreement No 101134999 (SOLER). The research was sponsored by the DynaSun project and has thus received funding under the Horizon Europe programme of the European Union under grant agreement (no. 101131534). Views and opinions expressed are however those of the author(s) only and do not necessarily reflect those of the European Union and therefore the European Union cannot be held responsible for them. SDO data are courtesy of NASA/SDO and the AIA. AW and EK acknowledge the SolMAG project (ERC-COG 724391) funded by the European Research Council (ERC) in the framework of the Horizon 2020 Research and Innovation Programme. AW, EK and JP acknowledge the Research Council of Finland Centre of Excellence project SpaceResilience (Grant number 374096). JP additionally acknowledges RCF project ENERGIZE (364852 and 370793).
\end{acknowledgements}

\bibliographystyle{aa}
\bibliography{Bibfile}

\begin{appendix}
\onecolumn

\section{Coronal dimming detection in all wavelengths}
\begin{figure*}[h!]
     \centering
     \includegraphics[width=0.9\linewidth]{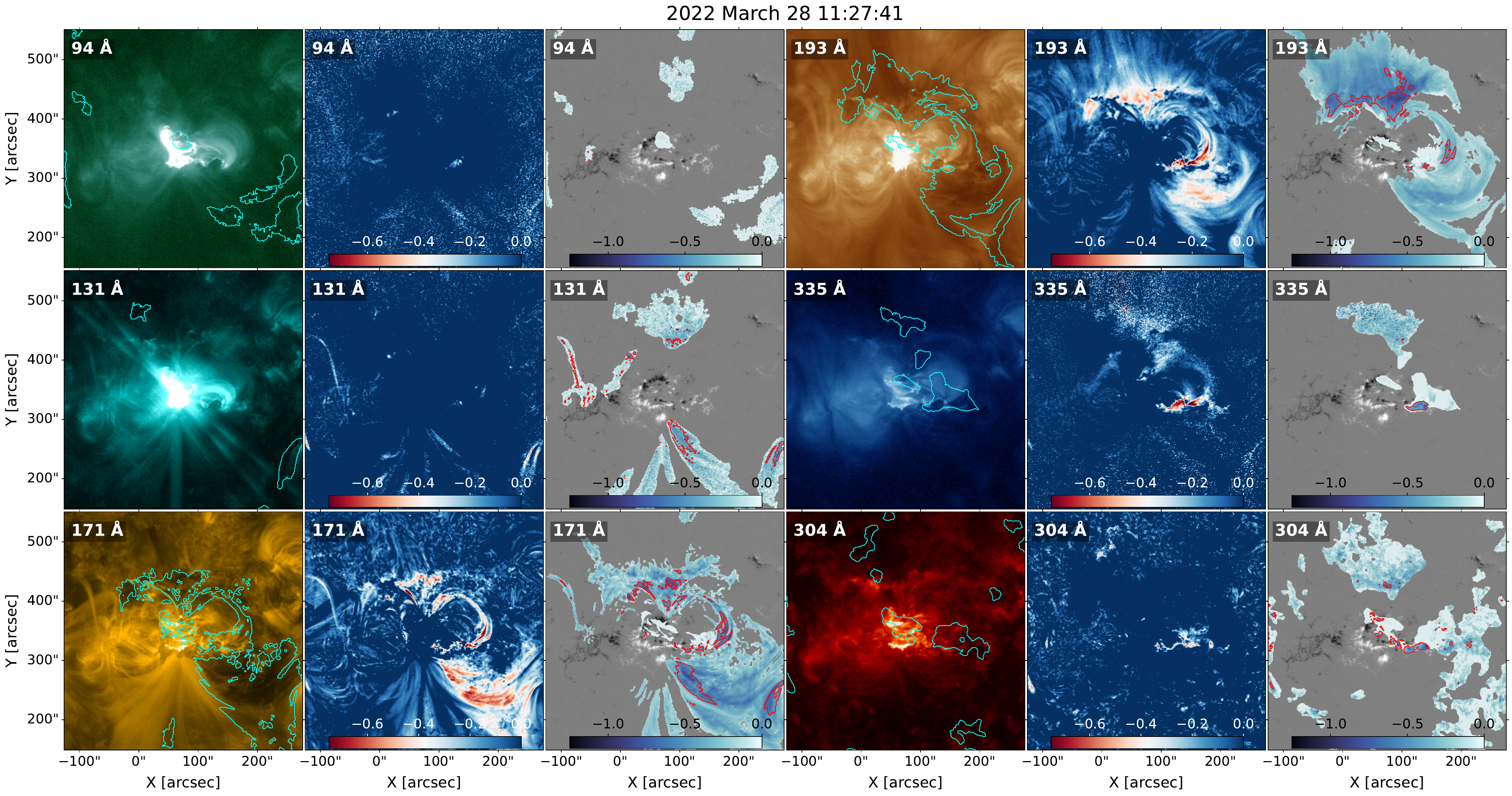}
     \caption{Evolution of the coronal dimming associated with the M4 flare on 28 March 2022. Similar to Fig.~\ref{fig:overview_dimming} but for the SDO/AIA 94, 131, 171, 193, 335, and 304~\AA~channels. For each wavelength, left panel: Direct image at 11:27~UT, with the dimming region detected up to that point contoured in cyan. Centre panel: Corresponding logarithmic base-ratio image. Right panel: Minimum intensity map from logarithmic base-ratio data overlaid on an SDO/HMI radial magnetogram, where the red outlines indicate core dimming regions detected with the method by \citet{Dissauer2018a}. The associated movie is available online.}
     \label{fig:overview_dimming_all}
\end{figure*}

We performed the coronal dimming detection method described in Sect.~\ref{sect:obs:dimming} for all EUV channels in SDO/AIA; that is, 94, 131, 171, 193, 211, 335, and 304~\AA.  Fig.~\ref{fig:overview_dimming_all} shows the dimming detection applied to the 94, 131, 171, 193, 335, and 304~\AA~channels, equivalent to Fig.~\ref{fig:overview_dimming} where the dimming detection method developed by \citet{Dissauer2018a} is applied to the AIA 211~\AA~channel. For each wavelength, we show a direct image of the event at 11:27~UT with cyan contours showing the dimming region detected up to that time step; a logarithmic base-ratio equivalent image at the same time step; and the final dimming region shaded on top of an SDO/HMI radial magnetogram. The colour of the shading shows the minimum intensity in logarithmic base-ratio units of each pixel during the detection time. An accompanying movie shows the temporal evolution of this figure.

The hotter AIA channels, 94 and 131~\AA, do not capture the southwestern footpoint dimming region, as no significant dimming is detected around the region. This is likely due to the absence of hot coronal loops above the southern footpoint, which is a sign of the complex configuration of this AR, thus resulting in low pre-event intensities in these channels. In contrast, the remaining wavelength channels are able to capture a dimming south of the AR, with the extent of the dimming being notably larger in the 171 and 193~\AA~channels than for the 335 and 304~\AA~channels. The southern footpoint region is successfully identified in these four channels using the method by \citet{Dissauer2018a}, and it is shown as red contours over the minimum intensity maps. The 335~\AA~channel is particularly effective at distinguishing the footpoint from the overall dimming region. As discussed in Sect.~\ref{sect:obs:dimming}, the northeastern flux rope footpoint is not detected in most wavelengths, and when it is, the signature is faint and blended with the erupting filament structure.

\end{appendix}
\end{document}